\documentclass[10pt, conference]{IEEEtran}
\IEEEoverridecommandlockouts
\usepackage[table]{xcolor}
\usepackage{multirow}
\usepackage{booktabs} 
\usepackage{amsmath}
\usepackage{graphicx}
\usepackage[belowskip=-12pt]{caption}
\usepackage{rotating}
\usepackage{tabularx}
\usepackage{subfig}
\usepackage{wrapfig}
\usepackage{listings}
\usepackage{color, colortbl}
\usepackage{balance} 
\usepackage{lscape}
\usepackage{xspace}
\usepackage{framed}
\usepackage{syntax}
\usepackage{parcolumns}
\usepackage{soul}
\usepackage[tikz]{bclogo}
\usepackage{enumitem}
\usepackage{makecell}
\definecolor{codegreen}{rgb}{0,0.6,0}
\definecolor{codegray}{rgb}{0.5,0.5,0.5}
\definecolor{codepurple}{rgb}{0.58,0,0.82}
\definecolor{backcolour}{rgb}{0.95,0.95,0.92}
\definecolor{redColor}{RGB}{255,0,0}
\definecolor{Gray}{gray}{0.1}
\lstdefinestyle{mystyle}{
	backgroundcolor=\color{backcolour},   
	commentstyle=\color{codegreen},
	keywordstyle=\color{magenta},
	numberstyle=\tiny\color{codegray},
	stringstyle=\color{codepurple},
	basicstyle=\scriptsize,
	breakatwhitespace=false,         
	breaklines=true,                 
	captionpos=b,                    
	keepspaces=true,                 
	numbers=left,                    
	numbersep=5pt,                  
	showspaces=false,                
	showstringspaces=false,
	showtabs=false,                  
	tabsize=2
}

\definecolor{diffstart}{named}{codegreen}
\definecolor{diffincl}{named}{redColor}

\lstdefinelanguage{CPlusPlus}{%
	language     = C++,
	morekeywords = {to_categorical, flow_from_directory,pad_sequences,load_image},
	morecomment=[f][\color{diffstart}]{+\ },
	morecomment=[f][\color{diffincl}]{-\ }
}
\lstdefinelanguage{Pythonna}{%
	language     = python,
	morekeywords = {to_categorical, flow_from_directory,pad_sequences,load_image},
	morecomment=[f][\color{diffstart}]{@@},
	morecomment=[f][\color{codegreen}]{+\ },
	morecomment=[f][\color{redColor}]{-\ }	
}
\lstdefinelanguage{LambdaCustom}{%
	language     = Lambda,
	backgroundcolor=\color{white},
	rulecolor=\color{black},
	identifierstyle=\color{black},
	frame=none
}
\lstdefinestyle{customc}{
	belowcaptionskip=1\baselineskip,
	breaklines=false,
	frame= single,
	breaklines = true,
	xleftmargin=\parindent,
	language= Pythonna,
	showstringspaces=false,
	basicstyle=\footnotesize\ttfamily,
	keywordstyle=\bfseries\color{green!40!black},
	commentstyle=\itshape\color{purple!40!black},
	identifierstyle=\color{blue},
	stringstyle=\color{codegreen},
	backgroundcolor=\color{gray!4}
}
\usepackage{caption}
\DeclareCaptionFormat{listing}{#3}
\graphicspath{{./figures/}}

\newcommand{\edge}{\textit{Microsoft Edge}}

\newcommand{\etal}{{\em et al.}\xspace}

\newcommand{\code}[1]{{\texttt{#1}\xspace}}

\newcounter{rqs}
\stepcounter{rqs}

\newtheorem{example}{Example}[section]

\newcounter{NumObservations}
\stepcounter{NumObservations}
\definecolor{shadecolor}{rgb}{.9,.9,.9}

\definecolor{msftBlue}{RGB}{0,164,239}
\definecolor{msftGreen}{RGB}{127,186,0}
\definecolor{msftYello}{RGB}{255,185,0}

\newcommand{\finding}[1]{
	\begin{bclogo}[couleur= msftBlue!60, epBord= 1, arrondi=0.1, logo=\bclampe,marge= 2, ombre=true, blur, couleurBord=msftBlue, tailleOndu=3, sousTitre ={\em #1}]{Finding \arabic{NumObservations} $\Rightarrow$ } 
		
	\end{bclogo}
	\stepcounter{NumObservations}
}

\newcommand{\ignore}[1]{{}}

\newcommand{\ourpara}[1]{%
\vspace*{3pt}%
\noindent\textbf{\textit{#1}}
}%

\usepackage[switch]{lineno}
\usepackage{cite}
\usepackage{balance}
\begin{document}
\title{Can Program Synthesis be Used to Learn Merge Conflict Resolutions? An Empirical Analysis \thanks{\noindent\rule{8.6cm}{0.1pt}
This work has been done when Nachiappan Nagappan was an employee at Microsoft and Rangeet Pan was an intern at Microsoft.}}
\thispagestyle{plain}
\pagestyle{plain}

\author{
    \IEEEauthorblockN{Rangeet Pan\IEEEauthorrefmark{1}, Vu Le\IEEEauthorrefmark{2}, Nachiappan Nagappan\IEEEauthorrefmark{2}, Sumit Gulwani\IEEEauthorrefmark{2}, Shuvendu Lahiri\IEEEauthorrefmark{2}, Mike Kaufman\IEEEauthorrefmark{2}}
    \IEEEauthorblockA{\IEEEauthorrefmark{1}Iowa State University, Ames, IA, USA
    \\rangeet@iastate.edu}
    \IEEEauthorblockA{\IEEEauthorrefmark{2}Microsoft Corporation, Redmond, WA, USA
    \\\{levu, sumitg, shuvendu.lahiri, mike.kaufman\}@microsoft.com, nnagappan@acm.com}
}
\maketitle
\begin{abstract}

Forking structure is widespread in the open-source repositories and that causes a significant number of merge conflicts.  
In this paper, we study the problem of textual merge conflicts from the perspective of \edge{}, a large, highly collaborative fork of the main \textit{Chromium} branch with significant merge conflicts. 
Broadly, this study is divided into two sections. First, we empirically evaluate textual merge conflicts in \edge{} and classify them based on the type of files, location of conflicts in a file, and the size of conflicts.
We found that $\sim$28\% of the merge conflicts are 1-2 line changes, and many resolutions have frequent patterns. 
Second, driven by these findings, we explore \textit{Program Synthesis} (for the first time) to learn patterns and resolve structural merge conflicts. 
We propose a novel domain-specific language (DSL) that captures many of the repetitive merge conflict resolution patterns and learn resolution strategies as programs in this DSL from example resolutions. We found that the learned strategies can resolve 11.4\% of the conflicts ($\sim$41\% of 1-2 line changes) that arise in the  C++ files with 93.2\% accuracy.
\end{abstract}
\section{Introduction}
\label{sec:introduction}

Textual merge conflicts occur when changes from independent commits over a common base cannot be resolved by a textual 3-way differencing tool present in version control systems such as git.
Merge conflicts are frequent and annoying, and studies have shown they can have a significant impact on the development lifecycle of projects~\cite{ghiotto-tse18}.
Prior works have focused on predicting merge conflicts~\cite{buse2010automatically, rastkar2013did, cortes2014automatically, fluri2007change, jiang2017automatically, owhadi2019predicting}, resolving them through using structured abstract syntax tree (AST) aware techniques~\cite{Apel11,lessenich2015,matsumoto2019beyond, raghavan2004dex,reuling2019automated}, or even verifying semantic correctness of merges using program verifiers~\cite{sousa2018verified}.
Although these approaches have paved the course for research on merge conflict prediction and resolution, the problem of merge-conflicts persists in practice. 
The reason can be attributed to (a) lack of structure-aware tools for most languages apart from Java~\cite{Apel11}, and (b) the black box nature of these tools that rely on sophisticated matching algorithms to relate the changes make them inaccessible to mainstream developers.

In this paper, we explore the problem of merge conflict resolution through a slightly different perspective. 
Instead of devising an algorithm to resolve any merge conflict, we pose the merge conflict resolution problem as that of {\it learning repetitive resolution patterns} from the history of a given project.
In fact, Ghitto \etal \cite{ghiotto-tse18} observed patterns are commonplace in conflict resolution for a large project when viewed over the history of a project.
They also postulate that learning common resolution patterns in a project can be a useful aid for developers in resolving future conflicts.


With this motivation, we conduct a large-scale empirical study of textual merge conflicts that occur in \edge{} (a fork of the \textit{Chromium} browser), and apply program synthesis (in particular, programming by example (PBE)~\cite{flashmeta,miltner2019fly}) to illustrate the feasibility of learning patterns for a large class of resolutions. 
We chose \edge{} as a subject due to (a) the collaborative nature of the project with 500+ developers and (b) the large number of merge conflicts that manifest as a result of merging changes from the \textit{Chromium} repository to this fork. 
The study consists of two main parts. First, we perform an empirical study to classify the nature of merge conflicts for this project in terms of the prevalence of file types, size of conflicts, type of conflicts, and the resolution patterns.
Second, we design a domain-specific language (DSL) to capture the resolution patterns in this project and use program synthesis to automate the learning of resolution strategies as programs in this DSL. 



In our rigorous empirical study of \edge{}, we found that a majority ($ \sim$47\%) of the files with conflicts were written in C++, and a significant number of changes ($\sim$28\%) are of 1-2 lines. 
Among these conflicts in C++, we find that 12.34\% are related to headers and macros.
More importantly, we observe that these small conflicts usually follow a few distinct resolution patterns. 
Specifically, we found 9 different patterns in all of C++ 1-2 liner changes and out of which, 6 patterns were found in \texttt{Include} and \texttt{Macro} related conflicts. 

\begin{figure*}
\centering
\noindent\begin{minipage}{.494\linewidth}
\scriptsize
\begin{lstlisting}[language = CPlusPlus,captionpos=b,caption=(a) Header Location Update Example 1, basicstyle=\scriptsize]
Forked Branch:
+ #include "ui/base/anonymous_ui_base_features.h"
+ #include "ui/base/mojom/cursor_type.mojom-shared.h"
Main Branch:
- #include "ui/base/cursor/mojom/cursor_type.mojom-shared.h"
\end{lstlisting}
\end{minipage}\hfill
\begin{minipage}{.488\linewidth}
\scriptsize
\begin{lstlisting}[language = CPlusPlus, captionpos=b,caption=(b) Header Location Update Example 2, basicstyle=\scriptsize]
Forked Branch:
+ #include "ui/base/anonymous_ui_base_features.h"
+ #include "ui/base/mojom/cursor_type.mojom-blink.h"
Main Branch:
- #include "ui/base/cursor/mojom/cursor_type.mojom-blink.h"
\end{lstlisting}
\end{minipage}
\end{figure*}
\begin{figure*}
\centering
\vspace{-29pt}
\noindent\begin{minipage}{.494\linewidth}
\scriptsize
\begin{lstlisting}[language = CPlusPlus, captionpos=b,caption=(c) Removal of a Specific Header Example 1, basicstyle=\scriptsize]
Forked Branch:
- #include "base/logging.h"
+ #include "base/scoped_native_library.h"
Main Branch:
+ #include "base/notreached.h"
\end{lstlisting}
\end{minipage}\hfill
\begin{minipage}{.482\linewidth}
\scriptsize
\begin{lstlisting}[language = CPlusPlus, captionpos=b,caption=(d) Removal of a Specific Header Example 2, basicstyle=\scriptsize, framexrightmargin=4pt]
Forked Branch:
+ #include "base/command_line.h"
- #include "base/logging.h"
Main Branch:
+ #include "base/check_op.h"
\end{lstlisting}
\end{minipage}
\vspace{-14pt}
\caption{Examples of Programming by Example (PBE) to Resolve Merge Conflicts in \edge.}
\label{fig:intro}
\end{figure*}
We describe one such pattern 
in Figure~\ref{fig:intro}(a), where the header ``cursor\_type.mojom-shared.h'' appears both in the forked and main branches and the developer excluded the one in the main branch (\textcolor{green}{\textbf{+}} and \textcolor{red}{\textbf{-}} denote whether the statement has been taken into the resolved solution or not. Only the changes with \textcolor{green}{\textbf{+}} operators are carried on to the resolution, whereas the changes with \textcolor{red}{\textbf{-}} operators are not).
Figure~\ref{fig:intro}(b) follows the same pattern, with the only difference being the specific name of the header file. 
Similarly, figures~\ref{fig:intro}(c,d) show another pattern, where the developer removes ``base/logging.h'' from the forked branch because \edge{} uses a different logging system.
These examples demonstrate the existence of common resolution strategies in the history of a project and the value of learning such frequent patterns to automate future conflicts.


Motivated by these patterns, we propose a domain-specific language (DSL) that contains the right set of abstractions to {\em succinctly} describe the patterns underlying the conflict-resolution examples. The key idea of our DSL is to describe the resolution as a guarded concatenation of fragments of the main and fork branches. The fragments are constructed using some ordered set operators including indexing, filtering, and subtraction.

Our DSL is able to express 11.44\% of conflict resolutions, which also accounts for $\sim$41\% of 1-2 line changes, seen in C++ files for \edge{} with 93.2\% accuracy.
Expressing the resolution strategies as programs can address the limitations of fully automated tools in the hands of mainstream developers as: (1) developers can examine the program and understand the justification behind suggesting the changes, and (2) if new patterns emerge and existing patterns do not work, a new program can be learned by either providing a few examples or updating the synthesized program. 

Our DSL, which contains the right set of abstractions that allow succinct expression of conflict resolution patterns, facilitates the development of a program synthesizer that can automatically learn intended patterns in this DSL from few examples. The key idea behind our synthesizer is to leverage a top-down synthesis methodology that reduces the synthesis problem to authoring efficient inverse functions for the various operators in our DSL~\cite{flashmeta}.

\subsubsection*{Contributions}
In summary, the paper makes the following contributions:
\begin{enumerate}[label=\alph*)]
    \item We empirically analyze textual merge conflict-related data from \edge{} to qualify various measures, e.g., which files are most prevalent to have conflicts, how big these conflicts are, and where the conflicts lie in a file. Finally, we systematically discover and establish frequent patterns in resolving merge conflicts.
    \item We build a domain-specific language (DSL) to address the \texttt{Include} and \texttt{Macro} related patterns established by the empirical evaluation.
    \item We leverage a top-down program synthesizer to learn the programs in this DSL to resolve the conflicts using input-output examples. 
\end{enumerate}

\subsubsection*{Organization} The paper is structured as follows: we discuss the dataset collection steps in \S\ref{sec:data} and the empirical study in \S\ref{sec:classification}. Then, we describe the domain specific language and the use of program synthesis for learning patterns through examples in \S\ref{sec:casestudy}. The results and related works are discussed in \S\ref{sec:result} and \S\ref{sec:related}, respectively. Finally, \S\ref{sec:conslucion} concludes.



\section{Dataset}
\label{sec:data}
\subsection{Main/Fork Branch Structure}
\label{subsec:structure}
\begin{figure}
    \centering
    \includegraphics[ 
    width=\linewidth]{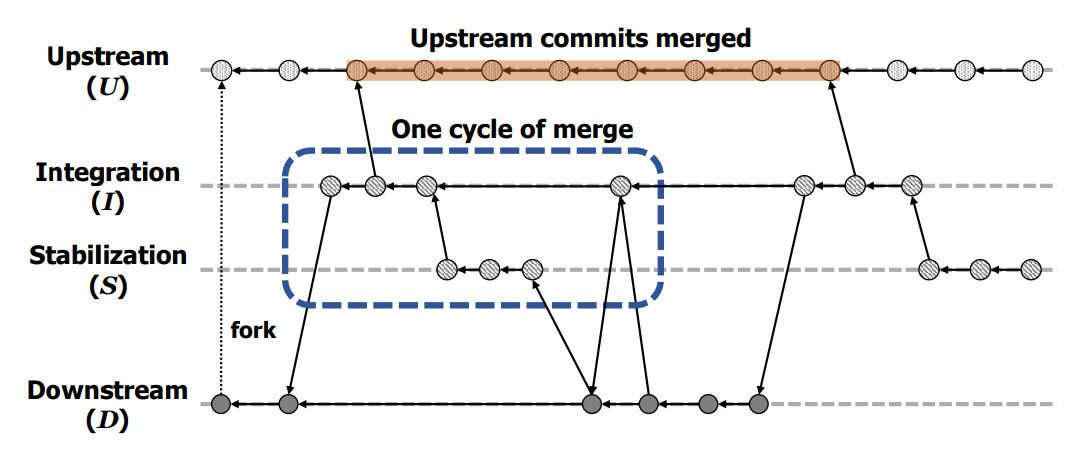}
    \caption{The Structure of the \edge{} Branch.}
    \label{fig:edge}
\end{figure}
\begin{table*}
\centering
\caption{Merge Related Data for \edge. M: Main Branch, F: Forked Branch}
\label{tb:summary}
\begin{tabular}{|l|c|c|c|c|c|c|c|c|c|c|c|}
\hline
\rowcolor{gray!35}
\multicolumn{1}{|c|}{Merge Related Data} & Week 1     & Week 2 & Week 3 & Week 4 & Week 5 & Week 6 & Week 7 & Week 8 & \multicolumn{1}{|c|}{Total}& \multicolumn{1}{|c|}{Median} & \multicolumn{1}{|c|}{SD} \\
\cline{2-9}
\rowcolor{gray!35}
\multicolumn{1}{|c|}{}                                    & 03/30- 4/5 &  4/6-4/12      &  4/13-4/19      &   4/20-4/26     &    4/27-5/3    &   5/4-5/10     &  5/11-5/17      &    5/18-5/24   & \multicolumn{1}{|c|}{} & \multicolumn{1}{|c|}{} & \multicolumn{1}{|c|}{} \\
\hline
\hline
Merges                                            &      37      &    45    &   43     &   32     &   37     &    29    &    76    & 38 &337  & 37.5  & 14.6   \\
\rowcolor{gray!25}
Commits in M                                 &      2031      &  1727    &    1835    &    2154    &    2068    &    2037    &    2213    & 1690     &15755& 2034.0   & 195.1   \\
Commits in F                    &      150      &     102   &   75     &  61      &    70    &   62     &    136    &  73 & 729 &   74.0 &  34.6  \\
\rowcolor{gray!25}
Conflicting Commits                           &   27    &       45     &   38     &     30   &    36    &    29    &  33      &   33 & 271  & 33.0  & 5.8  \\
Conflicting Files                                 &     214       &   269     &    225    &  272      &   289     &   276     &     353   &   165& 2063 & 270.5  & 56.4\\
\rowcolor{gray!25}
Conflicting Chunks                                 &    756        &    1212    &  533      &    529    &    5855    &    948    &     976   &  489 & 11298  &   852.0  & 1813.4\\
\hline
\end{tabular}
\end{table*}

Figure \ref{fig:edge} depicts the overall structure of \edge, where each line denotes a branch. The \textit{Main (M)} is the \textit{Chromium}'s master branch and \textit{Fork (F)} is the fork created by \edge{} to pull the changes from the \textit{M} to branch \textit{F}. The arrow ($\longrightarrow$) depicts the child-parent (fork-main) relation between two branches. For example, the arrow from the main (M) to fork (F) denotes that F is a fork of M. There are two more branches: \textit{Integration (I)} and \textit{Stabilization (S)}. These branches are intermediate checkpoints while pulling the changes from M branch. In each cycle, the changes from the \textit{M} branch are pulled to \textit{F} branch through \textit{I} and \textit{S} branches. First, the conflicts are resolved at the \textit{I} branch after merging with the \textit{M} branch. Once conflicts are resolved, a fresh build is performed at the \textit{S} branch and the changes in the \textit{S} branch are then pulled to the master branch of \edge{} (\textit{F} branch).

A similar main-fork branch structure can be found in other projects (Brave, Colibri, Epic, Samsung Internet, LG WebOS (TV) web engine, etc.), mobile applications (android fork), database (MariaDB), etc. 
In GitHub, there are more than 1.5 million forked projects in C++ alone \cite{git}.

\subsection{Data Collection}
First, we identify the merge commits by filtering the commits with two parents, one from the \textit{Main (M)} branch and one from the \textit{Forked (F)} branch. Next, we replay the merge commits for obtaining the conflicting files for each merge commit. Along with the conflicting file, we store the resolved version of each conflicting files.

In Table \ref{tb:summary}, the summary of the merge commits is shown. We collected the merge data for eight weeks (March 30 to April 24, 2020), and we divide the data by each week. In the first row, the total number of merge commits are shown. These merge commits denote the number of times \textit{F} branch pulled the changes from the \textit{M} branch. The second and third row denotes the total number of commits made by \textit{M} and \textit{F} branches during the period of time. The fourth row denotes the number of merges that result in introducing conflicts in at least one file. The fifth row shows the total number of files having at least one conflict. Finally, in the sixth row, we report that the number of conflicting chunks that have resulted from merging the two branches' changes.
We found that, on average, 80.4\% of the merges and 37.2\% of all commits at \textit{F} branch result in conflicts. Also, on average, $\sim$258 conflicting files are generated per week. We found that the number of merges (SD: 14.6), commits in main (SD: 195.1) and fork (SD: 34.6), conflicting commits (SD: 5.8), and conflicting files (SD: 56.4) to be very consistent over the 8 weeks. In comparison to the Java related conflicts~\cite{ghiotto-tse18}, \edge~(Mean 852) has bigger conflicting chunks than the ones found in the prior work (Mean 20). Based on the $p$-value and correlation value (corr), we found that conflicting chunk size ($p=0$, $corr=0.31$) has no impact on the number of conflicts.

\section{Classification of Merge Conflicts}
\label{sec:classification}
In this section, we empirically study the merge conflicts in \edge{} and answer four questions: 
\begin{enumerate}[label=\Alph*)]
    \item (\textbf{File Type}) Which file types are the most prevalent to have merge conflicts?
    \item (\textbf{Conflict Size}) How large are the conflicts?
    \item (\textbf{Conflict Location}) What are the different types of conflicts based on the origin of location?
    \item (\textbf{Resolution Pattern}) What are the conflict resolution patterns in \edge?
\end{enumerate}

\subsection{Which file types are the most prevalent to have merge conflicts?}
\begin{figure}
    \centering
    \includegraphics[trim={4cm 2cm 4cm 2cm}, width=\linewidth]{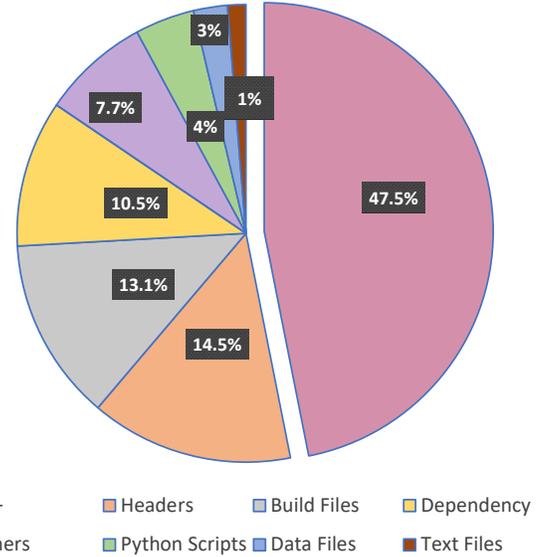}
    \caption{Classification Based on the File Type.}
    \label{fig:filetype}
\end{figure}
Here, we classify the conflicts based on the type of files. In Figure \ref{fig:filetype}, we denote the percentage of the conflicts for each file type. We found that most of the conflicts (47.5\% of the total conflicts) are in the C++ related files, and this is because the core functionality of both \textit{Chromium} and \edge{} are written in C++. In the following paragraphs, we briefly discuss the basic characteristics of each file type.
\paragraph{C++} The conflicts in these files are related to both logical and structural changes in the code. The structural changes include the addition, removal, or update of the header files, macros, etc. 
The logical changes are introduced to alter the logic of the operation. For example, addition of conditional check, loop structure, etc.

\paragraph{Dependency} These files are the dependency files that are needed to build \edge{} from scratch. Very often, dependencies, e.g., the version, controllable parameters, etc. are updated by the main and the forked branch. We found that various internal tools are used to automate the resolution of conflicts in this genre of files.

\paragraph{Headers} The definition and structures of the class are defined that are used in other files. 14.5\% of the files having at least one conflicts are header related ones.
\paragraph{Build Files} With each build (including run-time compilation), a considerable number of files are auto-generated. When they are pushed along with other changes, they often result in having conflicts. Since they are auto-generated codes, rebuilding can resolve the conflicts, and they do not need any particular logical or structural changes for resolution.
\paragraph{Python Scripts} 4.3\% of the conflicting files occur in the python scripts (.py and .pyl files). These files are utilized for running different scripting level codes.
\paragraph{Data Files} Mostly, these are matrix market files (.mm), and grid data format (.grd) files and changes in the data often cause the conflicts. However, only 2.4\% of the conflicts occur in these files.
\paragraph{Text Files} Changes in the documentation related files such as \textit{readme} files are categorized here.
\paragraph{Others} File types responsible for less than $1\%$ of the conflicting files are grouped in this category. Mostly, comma-separated files (.csv), protobuf files representing the structure of the code (.proto), etc. are in this group.
\subsection{How big are the conflicts?}
\begin{figure}
    \centering
    \includegraphics[trim={2cm 8cm 2cm 8.5cm}, width=\linewidth]{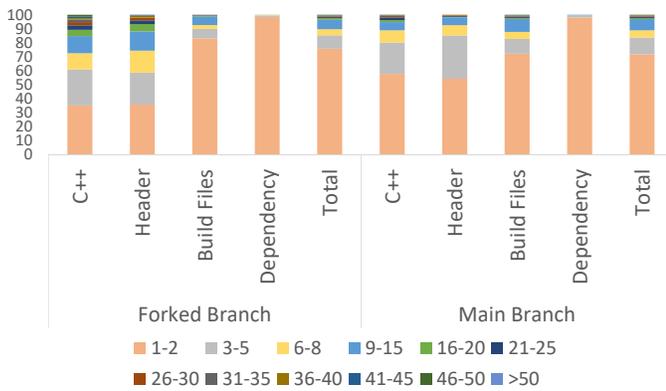}
    \caption{Classification Based on the Conflict Size.}
    \label{fig:size}
\end{figure}
\finding{35.17\% and 57.74\% of the conflicts in C++ are 1-2 lines of change by the main and forked branches, respectively. Overall, 28\% of the conflicts are of 1-2 lines for both main and forked branch.}

Each conflicting chunk has been divided into main and fork section based on the conflict markers. Furthermore, we categorize each section based on the number of lines of changes. We created twelve groups starting from 1-2 lines to more than 50 lines and compute the same for each type of files other than the ``others" category. Here, we refer to the conflict size as the number of lines reported by the conflict markers (main/fork section). We found that the majority of the conflicts are of 1-2 lines for either main (71.70\%) or fork (75.72\%). For C++ files, where the majority of the conflicts occur, we found that these numbers are 35.17\% and 57.74\% for main and fork, respectively.  Also, we found that 28\% of the conflicts in C++ are of 1-2 lines for both main and forked branches. Similar to what Ghitto \etal \cite{ghiotto-tse18} observed in Java projects, we found that a non-trivial amount of conflicts are due to small changes ($\le$8 lines) in C++ ($\sim$85\% and $\sim$75\% for main and fork, respectively). The distribution of the conflicts is similar to the one found in Java. In Java, 96\% of the conflicts are of $<$50 lines of code. Whereas in \edge, 98\% of the conflicts are of $<$50 lines of code. Furthermore, we computed the $p$-value (0) and the correlation value (0.35) and found that lines of code has no impact on the number of conflicts.

Given that smaller resolutions are more likely to be idiomatic and repetitive, we focus on the 1-2 lines of conflicts in both the main and forked regions. Moreover, since the majority of conflicts are in C++ files, we focus our study on the C++ files.

\subsection{What are the different types of conflicts based on the origin of location?}

\begin{figure}
    \centering
    \includegraphics[trim={0cm 0cm 0cm 0cm}, width=\linewidth]{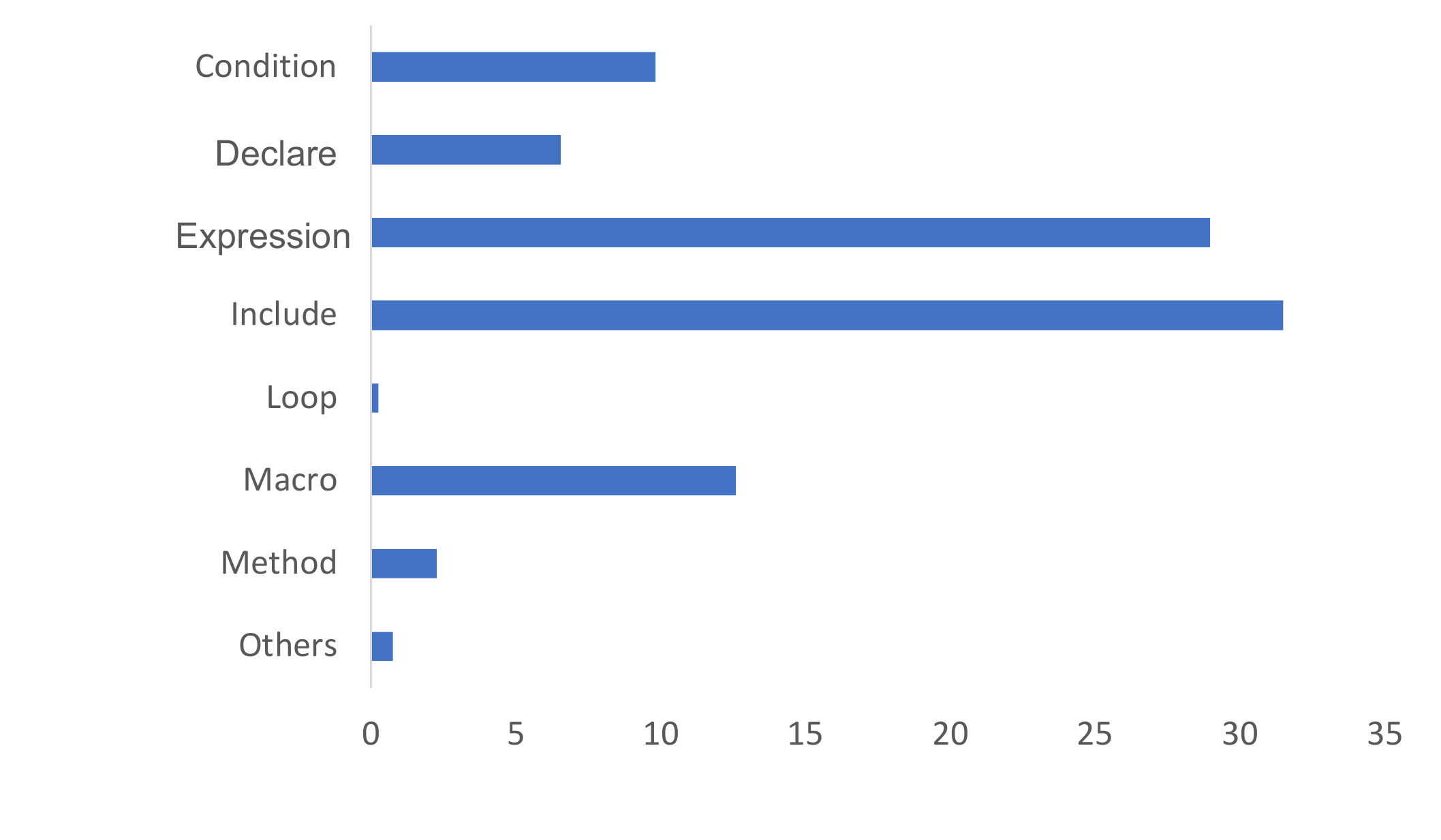}
    \caption{Classification Based on the Location of the Conflicts.}
    \label{fig:location}
\end{figure}
Based on the evidence that suggests the majority of the conflicts are in C++ files and are 1-2 lines in size, we move forward to classify the location of the conflicts. 

We utilized GumTree \cite{falleri2014fine} to tokenize the code and use the abstract syntax tree (AST) to determine the changes. We found that the conflicts mostly occur in seven locations that have been depicted in Figure \ref{fig:location}. In the following paragraphs, we briefly discuss each type of such location with appropriate examples.

\paragraph{Condition}
In this type of change, the conditional logic is updated. For example, in the code snippet below, the main branch added an extra condition that resulted in a conflict during the merge. This kind of conflict is 9.82\% of the total observed conflicts. This is similar to the percentage of condition related conflicts found in Java (9\%) by the prior work~\cite{ghiotto-tse18}.  
\begin{lstlisting}[language = c++]
Forked Branch:
  if (result != DID_NOT_HANDLE)
Main Branch:
  if (elastic_overscroll_controller_ && result != DID_NOT_HANDLE)
\end{lstlisting}
\paragraph{Declare}
In this type of change, the variable declarations are altered (6.55\% of the observed conflicts). In the following example, the value of the variable \textit{kSettingsVersion} has been changed to 1 from 4 by the main branch, and the data type has also been updated. 
\begin{lstlisting}[language = c++]
Forked Branch:
  static const uint32_t kSettingsMagic = 'CPds';
  static const uint32_t kSettingsVersion = 4;
Main Branch:
  static constexpr uint32_t kSettingsMagic = 'CPds';
  static constexpr uint32_t kSettingsVersion = 1;
\end{lstlisting}
\paragraph{Expression}
28.9\% of the conflicts originated due to changes in expressions. Change in the parameter while calling a procedure, calling a different procedure, etc., are categorized in this classification section.
\finding{31.49\% of 1-2 lines of conflicts in C++ files are due to addition, deletion, or updating the \texttt{Include} section.}
\paragraph{Include}
These types of changes are the most prevalent ones in the observed dataset. Mostly, the changes include the addition, deletion, and updating of the location of the headers.
For instance, in the following example, the main and forked branch added a new header file that caused the conflict to occur.
\begin{lstlisting}[language = c++]
Forked Branch:
#include "chrome/browser/ui/views/accessibility/hc_with_theme_bubble_view.h"
Main Branch:
#include "chrome/browser/ui/views/accessibility/caption_bubble_controller_views.h"
\end{lstlisting}
We found that 31.49\% of conflicts of size 1-2 lines are related to the include statement. This category of conflicts is higher than the observation made by Ghitto~\etal~\cite{ghiotto-tse18}, where 6\% of the conflicts in Java are due to the import statement (similar to the include statement in C++). The prior work has analyzed all the conflict sizes, whereas our analysis is based on the 1-2 liner conflicts.
\paragraph{Loop}
Only a small amount of conflicts (0.25\%) result from the change in the looping structure.

\paragraph{Macro}
C++ programs make extensive use of macros to enable code reuse.
We found that 12.6\% of the conflicts in C++ that are of 1-2 lines are due to changes in the macro. One such example has been depicted in the code snippet, where value corresponds to an argument has been changed by
\begin{lstlisting}[language = c++]
Forked Branch:
IN_PROC_BROWSER_TEST_P(V4,ANONYMOUS_DISABLED_PERMANENT(MalwareWithWhitelist, 20395305)) {
Main Branch:
IN_PROC_BROWSER_TEST_F(V4) {
\end{lstlisting}
the forked branch. In \edge, macros are mostly used to disable the main branch's (\textit{Chromium}) tests that do not work in \edge{} for various reasons, e.g., flaky test, or main branch tests a feature that \edge{} does not ship, or related to debugging. However, a large majority of the macro conflicts are more precisely test-disablement.
\paragraph{Method}
2.3\% of the changes that cause conflicts, are related to the changes in the method declaration, e.g., the return type, return value, etc. This genre of conflicts is less in C++ than in Java (10\%).
\paragraph{Others} In this category, we classify any conflicts that occur other than the previously mentioned ones. Mostly, they are situated in the comment block, and only 0.8\% of the conflicts under observation have been caused in these places.
\subsection{What are the conflict resolution patterns in \edge?}
Next, we empirically evaluate every 1-2 lines of changes (both main and forked branch) and identify the patterns based on the type of operations that the developers did to resolve the merge conflict. In this context, any resolution that removes the conflict markers is treated as a conflict resolution. We used the open coding scheme to build the classification scheme. First, we label the 10\% of the merge conflicts and build the classification scheme. Then, if needed, we add the classification category and revise the labels. Furthermore, the classification scheme has been verified by the third and the fifth authors, who are experts in software engineering study and merge conflicts. Finally, when no new classification category has been added, we completed the labeling process. A total of nine types of operations are found in our dataset. In the following paragraphs, we discuss each type of merge conflict, resolving patterns, and understand the root cause of such patterns with proper examples. 
\begin{figure*}
    \centering
    \includegraphics[trim={2cm 5cm 2cm 5cm}, width=0.95\linewidth]{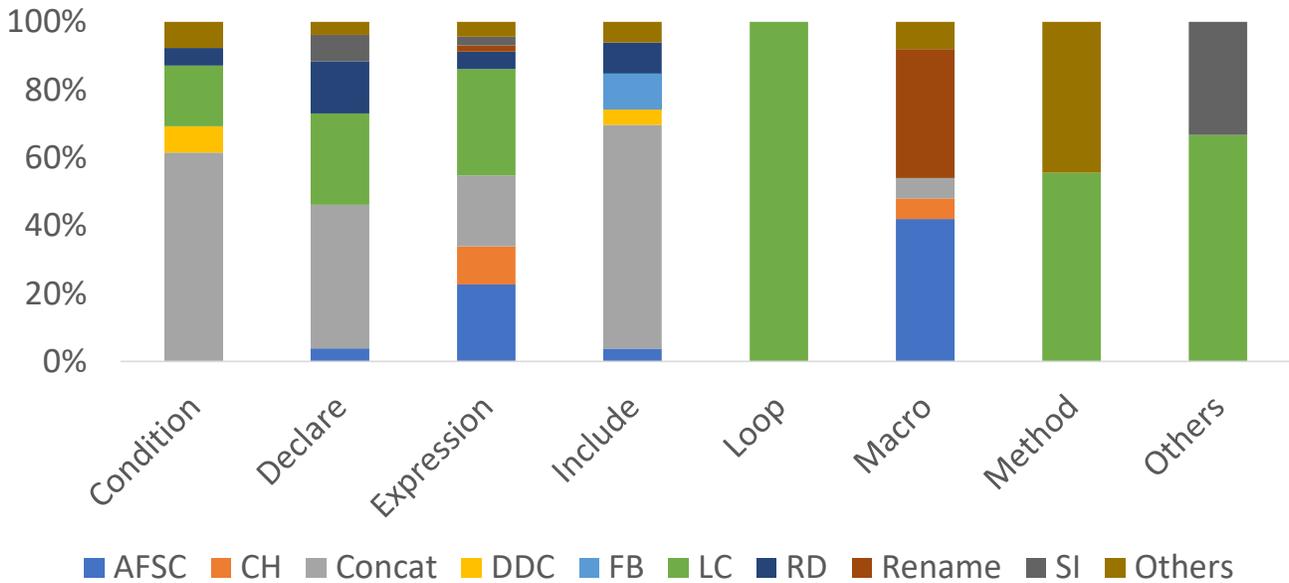}
    \caption{Classification Based on the Type of Operations that Developers Do to Resolve. \textbf{AFSC}: Apply Forked Specific Change, \textbf{CH}: Change in the Header, \textbf{Concat}: Concatenation, \textbf{DDC}: Dependency on different conflicts,	\textbf{FB}: Frequent Behavior, \textbf{LC}: Logical Change, RD: Remove Duplicate, and \textbf{SI}: Space Introduced.
}
    \label{fig:operation}
\end{figure*}
\paragraph{Apply Forked Specific Changes (AFSC)}
\edge{} has created a fork, and periodically they pull the changes from the Chromium to update the codebase. We found that developers from \edge{} update the code to reflect the \textit{Forked branch} specific changes, e.g., disabling tests, adding forked branch specific header, etc. When this kind of change is reflected in the merge conflict, the forked or \edge{} specific changes are taken into account, and the main branch-specific changes are discarded.
\begin{example}
\label{ex:afsc}
Apply fork specific changes, where \edge{} added a test disabling check.
\end{example}
\begin{lstlisting}[language = CPlusPlus]
Forked Branch:
+ IN_PROC_BROWSER_TEST_F(V4, ANONYMOUS_DISABLED_PERMANENT(CheckUnwantedSoftwareUrl, 20395305)) {
Main Branch:
- IN_PROC_BROWSER_TEST_F(V4, CheckUnwantedSoftwareUrl) {
\end{lstlisting}
In the code snippet, the forked branch introduced a check in the second parameter of the browser test case procedure, which has been used to disable a portion of the test cases in \edge. In the resolution of the merge conflict, the developer chose the fork branch changes discarding the main branch changes. These kinds of operations are mostly prevalent in changes located in macro (42.0\%), expression (22.6\%), and include statements (4.80\%).
\paragraph{Change in the Header (CH)}
A small portion of the conflicts is due to a change caused by another change in a different place (mostly in a header file).
For instance, the forked branch's change has added an extra parameter while calling the procedure. As the forked branch had made a change in the header (function \textit{OnExtensionDownloadFinished}) by adding a new parameter, the same change has been reflected in the C++ file to be consistent. While resolving the conflict, developers chose the latest changes (fork branch-related).
\begin{example}
\label{ex:ch}
A change in the header file caused the update of the following code.
\end{example}
\begin{lstlisting}[language = CPlusPlus]
Forked Branch:
+  ON_CALL(*this, OnExtensionDownloadFinished(_, _, _, _, _, _, _, _))
Main Branch:
-  ON_CALL(*this, OnExtensionDownloadFinished(_, _, _, _, _, _))
\end{lstlisting}

\paragraph{Concatenating Changes from two branches (Concat)}
\finding{39.5\% of the resolution strategies involved concatenating the main and the forked branch's changes.}
A vast majority of the resolution strategies are based on keeping both the changes that resulted in a conflict. For example, in case of the changes in the include statements, the developers can often take both changes without any particular order (if there is no commonality between the header files). Assume the main branch has made a change that involves the addition of the header \textit{header1.h}, and the forked branch added a header \textit{header2.h}. A developer often chooses to resolve the conflict by taking both the headers. Since these headers in \edge{} usually contain independent classes, the order in which the include statements are concatenated does not matter. However, for changes where more complex logic is involved, the concatenation requires to be more aligned with semantics rather than be on the syntactic changes. For example, in the code snippet, the resolution can be either the changes in the forked branch, followed by the changes from the main branch or the changes from the main branch, followed by the changes from the forked branch. If the latter strategy is taken, the header2 and header3 (actual name of the header has been changed) will be included in the conditional statement, so they would only be called if the conditional check returns true. However, this would create an anomaly in the code if there is a dependency on the header files irrespective of the value of the conditional check.
\begin{example}
\label{ex:concat}
Applying both the changes done by the forked and the main branch.
\end{example}
\begin{lstlisting}[language = CPlusPlus]
#if !defined(PROJECT_ANONYMOUS_BUILD)
#include <header1_name>
Forked Branch:
+ #endif
Main Branch:
+ #include <header2_name>
+ #include <header3_name>
\end{lstlisting}

\paragraph{Dependency of Different Conflicts (DDC)}
\label{para:ddc}
This type of resolution strategy accounts for a small number of scenarios (2.3\%). If more than one conflict is present in the C++ file and one decision of resolution affects the resolution of another conflict, we label the fixing strategies as this group. 
\begin{example}
\label{ex:dependency}
A header file has a dependency on a section of code, which is also in the conflicting region.
\end{example}
\begin{lstlisting}[language = CPlusPlus]
Forked Branch:
+ #include "components/project_anonymous/core/common/switches.h"
- #include "components/version_info/channel.h"
Main Branch:
\end{lstlisting}
The header file (\textit{``components/version_info/channel.h"}) has been referred to in a single place of the file, and that section was deleted while resolving another conflict. Here the reference is denoted by the class-object relation. The header file, \textit{``channel.h"}, has a class named \textit{channel} declared, and the object of that class has been used in the C++ file that includes the header statement. Since the object reference is no longer present, the developer decides to remove the \texttt{Include} statement as well.
\paragraph{Frequent Behavior (FB)}
\label{para:freq}
We found that there are some repetitive patterns, where developers remove a specific header file while fixing the merge conflicts. For example, 92.8\% of (13 out of 14) the scenarios involving the \textit{``base/logging.h"} header file, the line related to the specific header file has been deleted, keeping other changes intact (Figure \ref{fig:intro}(c) and (d)). This is a project-specific pattern, which is due to the fact that fork is using the logging function differently than the main.

\paragraph{Logical Changes (LC)}
This genre of fixing strategies needs a logical understanding of the code. In most cases, the resolve includes a section of the main and the forked branch stitched by an appropriate logic. 
\paragraph{Remove Duplicates (RD)}
These fixing strategies are mostly found in the condition (5.13\%), declare (15.38\%), expression (5.22\%), and include (4\%) related conflicts. Overall, 4.28\% of the conflicts under observation are resolved by removing the duplicate content.
Often, the main or forked branch moves a file into a different location and updates the same in the code. To resolve such conflicts, developers tend to remove the duplicate \texttt{Include} statement(s). For example, the code snippet in Figure \ref{fig:intro}(a) has the same header name \textit{``cursor\_type.mojom-shared.h"} in the main and the forked branch. However, the location of the file has been changed without changing the content of the header file. In such a scenario, the developer can choose any version of the \textit{``cursor\_type.mojom-shared.h"} file and take the rest of the changes from the main and the forked branch. The same can happen with header files outside the conflicting region. 

Another scenario includes the variable declaration. In the following scenario, the variable \textit{supervised_user_id} has not been initialized (main branch) or initialized with a default value (forked branch). However, in C++, the effect of either statement will result in the same effect. If the string is not initialized, then the default constructor will be called, and it will assign a zero size character to the string, which is similar to initializing with the empty string (\textit{""}). To resolve such conflict, the developer can choose any version of the variable declaration and add that with the \textit{\#endif} line.

\begin{lstlisting}[language = CPlusPlus]
Forked Branch:
+ #endif
+ std::string supervised_user_id = "";
Main Branch:
- std::string supervised_user_id;
\end{lstlisting}

\paragraph{Rename Related Fixes}
File renaming is very common in a fork structure. 
In our case, resolving the rename related operation accounts for 5.29\% of conflicts with 1-2 lines of change, and they are most prevalent in macros (38\%). In this scenario, any of the two parties involved have changed the name of the function, and to make it consistent, every place where the function has been called, has been changed too. This kind of change often results in conflicts, and one such instance has been depicted in the following example.
\begin{example}
\label{ex:rename}
\edge{} test has been renamed.
\end{example}
\begin{lstlisting}[language = CPlusPlus]
Forked Branch:
- IN_PROC_BROWSER_TEST_P(V4,                      ANONYMOUS_DISABLED_PERMANENT(CheckResourceUrl, 20395305)) {
Main Branch:
+ IN_PROC_BROWSER_TEST_F(V4, CheckResourceUrl) {
\end{lstlisting}
The name of the test has been updated by the main to \textit{IN_PROC_BROWSER_TEST_F} from \textit{IN_PROC_BROWSER_TEST_P}. In such cases, the developer chose the renamed version of the function name and resolved the conflict.
\paragraph{Space Removed (SR)}
Introducing an extra space in the code can cause a conflict. As the default merging strategy is based on the textual difference, these scenarios are also considered as a case for merge conflict. In such scenarios, developers remove the space.

\paragraph{Others} Here, we labeled the conflicts that have no pattern of fixing or not enough information to conclude.
\section{Case Study on Deploying Synthesis for Resolving Merge Conflicts}
\label{sec:casestudy}
The empirical study has identified several patterns for resolving merge conflicts. We used these high-level patterns to guide the DSL to resolve the conflicts.
We now discuss the use of program synthesis to learn programs expressing these patterns from examples. We first use some examples from the previous section to motivate the design of our DSL.
We then formally define the DSL, and discuss the learning of programs in this DSL from examples using PROSE, an inductive program synthesis framework~\cite{flashmeta, prose}. The DSL that we propose possesses some syntactic sugar to be able to convey the high-level patterns. We did not want to break the patterns into assembly-level languages.

\subsection{Motivating Examples}
\label{sec:dsl-examples}
Recall that in Figure~\ref{fig:intro}(c) and (d), the desired behavior is to remove the header ``base/logging.h" from the forked branch's changes, then concatenate the changes in both branches. 
Additionally, we need a \emph{guard} to only execute the resolution above when the forked branch contains ``base/logging.h". Otherwise, the resolution would blindly apply to other non-applicable cases. Here is a program that expresses the pattern above:
{
\small
\texttt{Apply}(\texttt{FrequentPattern}($x$, \text{``base/logging.h"}),\\
    \hspace*{10mm}\texttt{Concat}(\texttt{Main}($x$), \\
     \hspace*{12mm}\texttt{Remove}(\texttt{Fork}($x$), \texttt{ForkByPath}($x$, \text{``base/logging.h"})))
}%

In this program, \texttt{Apply} checks if the forked branch of the input $x$ contains the header ``base/logging.h" (using the predicate \texttt{FrequentPattern}), then returns the resolved text as the concatenation of the main branch and the forked branch having the header ``base/logging.h" removed.

In Figure~\ref{fig:intro}(a), the header ``cursor\_type.mojom-shared.h" appears in both the main and forked branches, hence only the header in the forked branch is selected. 
Although we can use a predicate to check if both main and fork contain that particular header, such a predicate is very specific to that header name and cannot capture other headers.
Instead, we introduce a predicate \texttt{DuplicateMainFork} to check if there are any headers that appear in both main and forked branches.
We then resolve the merge by concatenating (a) the forked branch and (b) the main branch with these duplicated headers removed (note that \texttt{Pattern(\dots)} returns the duplicated headers).
{\small
\texttt{Apply}(\texttt{DuplicateMainFork}($x$),\\
    \hspace*{10mm}\texttt{Concat}(\texttt{Fork}($x$), \\
    \hspace*{12mm} \texttt{Remove}(\texttt{Main}($x$), \texttt{Pattern}($x$, \text{``DuplicateMainFork"})))
}%
\subsection{DSL for Resolving Merge Conflicts}
\label{subsec:grammar}
\begin{figure}[h]
\small
  \begin{align*}
    \text{Output } {\bf r} := {}& \texttt{Apply}(\text{c}, \text{t})\\
    \text{Condition } {\bf c} := {}& \texttt{And}(\text{p}, \text{c}) \mid \text{p}\\
    \text{Predicate } {\bf p} := {}& \texttt{DuplicateMainFork}(x)\\
    \mid  {} &\texttt{DuplicateMainOutside}(x)\\    
    \mid  {} &\texttt{DuplicateForkOutside}(x)\\    
    \mid  {} &\texttt{MainSpecific}(x) \mid \texttt{ForkSpecific}(x)\\    
    \mid  {} &\texttt{Dependency}(x) \mid \texttt{Rename}(x)\\    
    \mid  {} &\texttt{FrequentPattern}(x, path)\\    
  \text{Transformation } {\bf t} := {}& \texttt{Concat}(\text{t}, \text{t}) \mid \texttt{Remove}(\text{s}, \text{s}) \mid \text{s}\\
    \text{Selection } {\bf s} := {}& \texttt{Main}(x) \mid \texttt{Fork}(x)\\
    \mid  {} &\texttt{ForkByPath}(x, path) \mid \texttt{ForkByIndex}(x, k)\\
    \mid  {} &\texttt{MainByIndex}(x, k) \mid \texttt{MainByPath}(x, path)\\
    \mid  {} &\texttt{Pattern}(x, key)
  \end{align*}
\caption{Syntax of our DSL. $x$ refers to the input. $path, k, key$ denote a path constant, an integer constant, and a string constant, respectively. }
\label{fig:dsl}
\end{figure}

Having demonstrated some program constructs in our DSL, we now discuss the DSL in detail. Figure~\ref{fig:dsl} shows its syntax.
At a high level, a program takes an input $x$, a struct that contains the main and fork sections denoted by the conflict marker, the location of the file, the rest of the content of the file, and returns the resolved text. We take the location of the file to extract the content of the imported header files and utilize them to detect the patterns, e.g., the example depicted in \S\ref{ex:dependency}, where we check the rest of the content and the content of the header files along with the content between the conflict markers. The program uses \texttt{Apply} to check if some conditions {\bf c} hold, then performs the merge resolution {\bf t} that transforms the common base program to a resolved program. 

\ourpara{Condition} The condition \textbf{c} is a conjunction of one or more predicates.
Our DSL has the following predicates to capture the classification discussed in Section~\ref{sec:classification}:

\begin{itemize}
    \item \texttt{DuplicateMainFork:} Some content appears in both main and forked branches. Figures~\ref{fig:intro} (a,b) illustrate this predicate, in which the header ``cursor\_type.mojom-shared.h" and ``cursor\_type.mojom-blink.h" have the same name and content in both main and fork.

    \item \texttt{Duplicate[Main/Fork]Outside:} Some content appears in both the main (or forked) branch's conflicting region and non-conflicting region. 
    
    \item \texttt{MainSpecific/ForkSpecific:} The main (or forked) branch has some identifiers that are main (or fork) specific. For instance, the change in Example~\ref{ex:afsc} is fork specific because it has the macro \textit{ANONYMOUS_DISABLED_PERMANENT}.
    
    \item \texttt{Dependency:} One or more headers have dependency (control dependency) on other conflicting regions. Example \ref{ex:dependency} illustrates this predicate, where ``channel.h" header has a dependency on another conflicting region.
    
    \item \texttt{FrequentPattern:} This predicate captures a repetitive pattern when a specific header (identified by the literal $path$) is present in the conflicting regions. Figure~\ref{fig:intro} (c) and (d) illustrate this predicate in which $path$ is ``base/logging.h".
    
    \item \texttt{Rename:} Check whether the name of the tests in the macro has been altered or not. Example~\ref{ex:rename} uses this predicate to capture rename operation.
\end{itemize}


\ourpara{Transformation} During the study, we observed that the merge resolution usually involves the concatenation of some nodes selected from the main or forked branches. 
The resolution also occasionally removes some nodes from a branch because they appear in both branches.
Based on this observation, we designed the transformation {\bf t} that allows arbitrary combination of concatenation (\texttt{Concat}) and removal (\texttt{Remove}) of nodes. 
The key insight of designing a DSL amenable for synthesis is that the DSL should be expressive enough to cover common patterns, yet it should be restrictive enough to allow for efficient learning.
In our DSL, while \texttt{Concat} allows concatenation of two other transformations {\bf t},
\texttt{Remove} only allows removal of some selected nodes {\bf s} from some other selected nodes {\bf s}. Restrictions like these reduce the search space for transformation significantly, while still allowing for the expression of all patterns in Section~\ref{sec:classification}.

Our DSL allows selecting the whole main or forked branch via \texttt{Main} and \texttt{Fork}.
It also supports selection using index or path (\texttt{[Main|Fork]By[Index|Path]}).
The former selects the $k^{th}$ node from a branch while the later selects a node based on path. 
Furthermore, some patterns in our study require selecting nodes that are relevant to the predicates. By selecting Main and Fork, we want to identify only the main related section of the conflict marker. The nodes are AST nodes. However, because our DSL only works with \#include and macros, we simplify the nodes to contain only relevant information. For instance, an include statement $\texttt{\#include <foo>}$ corresponds to a node that has two children, one for ``\#include'' and another one for $\texttt{<foo>}$.

We allow such selection via \texttt{Pattern}, where a string $key$ is used to identify the predicate name.
In the example in Section~\ref{sec:dsl-examples}, $\texttt{Pattern}(x, \text{``DuplicateMainFork"})$ selects nodes related to the predicate \texttt{DuplicateMainFork} (i.e., the nodes appearing in both the main and forked branches). Internally, our program calculates a dictionary that maps each pattern to some nodes.
The dictionary is used both in evaluating the predicate and selecting the pattern nodes.




\ignore{
In order to describe a synthesis problem, a domain-specific language $\mathcal{L}$ is needed. This language $\mathcal{L}$ can be thought of as a context-free grammar (CFG), and each symbol $\mathcal{S}$ described is a single rule or a combination of rules. Each such symbol $\mathcal{s}$ can further be expanded by defining all the possible operators. These operators are described on the right-hand side of the symbol, and both the output type of the symbol and operators should be the same. Each such operators can be either a terminal node in the CFG or can further be decomposed into sub-programs. These operators can take the input as well as the free-variables available at the scope. These free-variables can either be defined in the language or can be introduced using the \textit{let} operator. The need for the symbol in the language is to transform a program $\mathcal{P}$ from a state to another state and finally to a value that can no longer be decomposed.

In our DSL $\mathcal{L}$, we have input (denoted by \code{input} in the DSL), which is the content of the upstream merge conflict, downstream merge conflict, the content of the file, and the location of the repository. Since, we are working on the include and macro locations, we pre-process each line of the conflict into the path present in the include statement and the whole line for the macro after removing the spacial characters and the new line symbols. We consider each such path or line as a single \code{Node} and the output of the program would be a list of \code{Node}s. Our DSL starts from defining a set of rules, which are indeed a transformation of the non-terminal symbols into another non-terminal symbol, value, or a \code{let} function. In the next paragraph, we will discuss each such symbols and rules briefly.

\textbf{proc:} This is the starting point of the DSL, and it can only be decomposed into \code{rule}.

\textbf{rule:} This is a \code{let} definition, and it returns the list of nodes, which are the resolved part of the merge conflict. This \code{let} function will create a local variable \code{dup} that has been returned by the \code{find} and has been under the scope of operation in the \code{Apply}. The output of the \code{Apply} has been treated as the output for the \code{rule}.

\textbf{find:} This function will take the input and a list of nodes (the program will learn this list of nodes) and return a list of nodes based on the operations performed at \code{FindMatch}.

\textbf{Apply:} This rule will return the list of nodes, which is the resolved output of the merge conflict. It takes the \code{predicate}, which is a boolean variable, and we take action only when we match the pattern of the input with the learned pattern.

\textbf{predicate:} Each learned program will have a set of signatures that it matches with the new input. For example, we have $N$ number of patterns we support, and we create a list that actually tracks, which patterns are applicable for a program. We call that \code{enablePredicate}, where store the patterns and if the patterns match with the \code{dup}, which is the output of the \code{FindMatch}, then move forward to apply the action.
}

\subsection{Synthesizing Programs from Examples}
We now discuss the synthesis process to learn programs in our DSL from examples using PROSE~\cite{prose}. 
PROSE framework is a meta-synthesizer, i.e., a system that allows people to implement inductive synthesizers more easily by re-using shared components, which would otherwise have to be reimplemented in different synthesizers.
Prior to PROSE, authors of inductive synthesizers such as FlashFill~\cite{flashfill}, FlashExtract~\cite{flashextract} had to implement their synthesizers from scratch. Newer synthesizers such as Refazer~\cite{rolim2017learning}, BluePencil~\cite{miltner2019fly} leverage PROSE to speed up the development. We follow the same strategy.
To create a new synthesizer in PROSE, one needs to (1) design the DSL and its semantics, (2) implement the witness/learning functions, and (3) define a ranking scheme for programs. All other low-level details such as building the version-space algebra, intersection of program sets, are taken care of by the PROSE framework.
Readers may refer to~\cite{flashmeta} for a more detailed discussion on inductive program synthesis, as well as exemplar synthesizers on other domains.

The input (also called \emph{specification}) to the inductive synthesizer is some examples in the form of $\{\sigma_i, o_i\}_{i=1}^n$ where $\sigma_i$ is the input $x$, $o_i$ is the expected resolved output, n is the number of examples.
The PROSE methodology requires each DSL function to be accompanied with a \emph{witness function} (WF, also called inverse function), which takes the output and returns the set of all input configurations to that function that can generate that output. 
Notice that a WF does the \emph{inverse} of its normal function, which takes the inputs and returns the output.
PROSE uses the WFs to decompose the input specification into ``smaller'' specifications for each of the parameters, effectively decomposing a complicated learning task into several smaller, simpler tasks. The divide-and-conquer process continues until it reaches the leaf nodes in the grammar, at which point the leaf nodes are either variables or literals.

Given a specification $\{\sigma_i, o_i\}_{i=1}^n$, the WF of \texttt{Apply(c,t)} reduces it to: (a) $\{\sigma_i, \text{true}\}_{i=1}^n$ for condition \texttt{c}, where we need to learn a conjunction of predicates that returns true on inputs $\sigma_i$, and (b) $\{\sigma_i, o_i\}_{i=1}^n$ for transformation \texttt{t}, where we need to learn a transformation that converts inputs $\sigma_i$ to outputs $o_i$.

\ourpara{Learning Conditions}
The task here is to find predicates matching the patterns in the input $\sigma_i$. 
We simply select all predicates, whose evaluation on the input is \texttt{true}.
One exception is \texttt{FrequentPattern}, where we also need to synthesize all possible values for $path$ such that applying the predicate on $path$ returns \texttt{true}.


\ourpara{Learning Transformations}
The goal here is to learn a sequence of transformations that transforms $\sigma_i$ to $o_i$ for all $i$.
While learning \texttt{Concat}, its WF decomposes the output $o_i$ into
all pairs $o_{i1}$, $o_{i2}$ such that \texttt{Concat}($o_{i1}, o_{i2}$) = $o_i$. 
This yields two sub-tasks with specifications $\{\sigma_i, o_{i1}\}_{i=1}^n$ and $\{\sigma_i, o_{i2}\}_{i=1}^n$. If PROSE finds transformations \texttt{t1} and \texttt{t2} for these sub-tasks, it returns \texttt{Concat(t1, t2)} as the result of learning this function.
E.g., in Figure~\ref{fig:intro}(d) where the resolved output contains two includes for ``base/command\_line.h" and ``base/check\_op.h", \texttt{Concat}'s WF assigns each include to a sub-task. PROSE recursively finds if there is a transformation \texttt{t1} that produces ``base/command\_line.h", and \texttt{t2} that produces ``base/check\_op.h".

The learning of \texttt{Remove} is similar; its WF decomposes the output $o_i$ into 
all pairs $o_{i1}$, $o_{i2}$ such that \texttt{Remove}($o_{i1}, o_{i2}$) = $o_i$. In the above example, one way to obtain the include for ``base/command\_line.h" ($o_i)$ is to select the forked branch (which contains ``base/command\_line.h" and ``base/logging.h" - $o_{i1}$) and removes ``base/logging.h" ($o_{i2}$) from it.

The learning of \texttt{[Main|Fork]By[Index|Path]} functions simply returns the index or the path of the output in the (main|fork) input, depending on whether the function is index-based or path-based.
For instance, to produce ``base/check\_op.h", two possible selections are $\texttt{Main}(x)$, where we select the main branch, or $\texttt{MainByIndex}(x, 0)$, where we select the first line of the main branch.

To learn $key$ for \texttt{Pattern}, we iterate over all patterns and return those whose nodes match the output. E.g., in Figure~\ref{fig:intro}(a), the pattern DuplicateMainFork produces the duplicate node ``ui/base/cursor/mojom/cursor\_type.mojom-shared.h", hence the key ``DuplicateMainFork" is selected.

\ourpara{Ranking Scheme}
Since there are often multiple programs that satisfy a given set of examples, inductive synthesizers use a ranking scheme to select the most likely program. 
For instance, in Figure \ref{fig:intro}(c), here are two possible choices for transformation {\bf t} (among many others):\\
$\circ\;\;$\texttt{Concat(MainByIndex(x,0),ForkByIndex(x,0))}\\
$\circ$\texttt{ Concat(Main(x), Remove(Fork(x),} \\\hspace*{20mm}\texttt{ForkByPath(x,"base/logging.h")))}\\
Although both programs are consistent with this scenario, our ranking scheme selects the latter program, which happens to be the intended one. We defined our ranking function to be a small weighted combination of a set of features that we discovered were relevant for our domain. For instance, we prefer programs that are smaller and use fewer constants since such programs generalize better to unseen inputs; hence, our feature set includes the number of operators and the number of constants in the program. Another feature accounts for favoring path-based programs to index-based programs because the former is more general, or for favoring selection from one of the branches since that is more likely. 


\section{Results}
\label{sec:result}
We now present our experimental results of using program synthesis to automate merge conflict resolution.

Apart from the 8-week dataset that is used to design our DSL and synthesizer, we also collected merge conflicts for another 4 weeks (March 2 to March 29, 2020) to evaluate our approach.

In this section, we answer the following questions:
\begin{enumerate}
    \item {\bf RQ1:} What portion of the merge conflicts can be automated?
    \item {\bf RQ2:} How accurately can we assist developers in resolving the merge conflicts?
\end{enumerate}

\noindent\textit{\textbf{RQ1: What portion of the merge conflicts can be automated?}}
Our empirical study found that $\sim$28\% of the C++ conflicts are of 1-2 lines of changes for both the main and fork branch. Among those, \texttt{Include} and \texttt{Macro} related conflicts account for 31.5\% and 12.6\%, respectively. For \texttt{Macro} related conflicts, we focus on the test-disabling macros by selecting the conflicts having \edge{} specific keywords such as \texttt{DISABLED}, \texttt{ANONYMOUS} (removed for double-blind) in the argument. We found that $\sim$78\% of the \texttt{Macro} related conflicts are of test disabling. For \texttt{Include} related issues, 8 cases (mentioned as \texttt{Others} in the classification) had files deleted when the developers resolved the conflict. 
We omitted those scenarios from our dataset because we were not able to get the resolution. 
Overall, our approach handles 11.44\% of the total conflicts in C++ files.

\noindent\textit{\textbf{RQ2: How accurately can we assist developers while resolving the merge conflicts?}}
\finding{Overall, our approach handles 11.44\% of the total conflicts in C++ files, which is 40.9\% of all the 1-2 line changes in C++.}
\begin{table}
\centering
\caption{Results of Using Program Synthesis on \edge over 8 weeks data.}
\label{tb:result}
\begin{tabular}{|l|c|r|}
\hline
\rowcolor{gray!35}
Pattern Name & Pattern Count & User Resolution \\ \hline
\hline
Concat       & 88            & 83 (94.3\%)        \\ \hline
\rowcolor{gray!25}
AFSC (Include)        & 5             & 4 (80.0\%)         \\ \hline
RD           & 12            & 11 (91.67\%)       \\ \hline
\rowcolor{gray!25}
FB           & 14            & 13 (92.8\%)        \\ \hline
DDC          & 6             & 5 (80.0\%)         \\ \hline
\rowcolor{gray!25}
Rename       & 15            & 14 (93.33\%)        \\ \hline
AFSC (Macro)  & 22            & 21 (95.4\%)        \\ \hline
\rowcolor{gray!25}
Total  & 162            & 151 (93.2\%)        \\ \hline
\end{tabular}
\vspace{-8pt}
\end{table}
Table~\ref{tb:result} presents our results on various \texttt{Include} and \texttt{Macro} scenarios on our 12-week dataset.
In this table, the column \texttt{User Resolution} represents the total number of resolutions provided by our approach that match with the resolution performed by the developers. To perform the experiment, we extract the solutions after developers resolve all the conflicts in the commits and match that with our proposed resolution.
We found that overall, our programs can assist the developers to resolve the merge conflict with 93.2\% accuracy. If the conditions depicted in our DSL in \S\ref{subsec:grammar} do not match the type of merge conflict that appears in the new examples, then our approach will not suggest.
Also, these programs achieve 91.17\% accuracy on the unseen 4-week dataset that we collected after our initial study. 
Because the conflicts in the 4-week dataset are not classified, 
we did not report its detailed results by classification in Table~\ref{tb:result}.
Instead, we report the frequency of patterns detected while applying programs on this dataset. We found that Concatenation is applied 73.5\% of the cases, Apply for Fork Specific (both \texttt{Include} and \texttt{Macro}) 8.8\%, Frequent Behavior 5.88\%, and Remove Duplicate 11.8\%.
These results indicate that program synthesis can be applied to resolve merge conflicts. 


\section{Related Work}
\label{sec:related}
\ourpara{Merge Conflicts:}
In this paper, we studied the problem of repeated patterns in textual merge conflict resolution in large projects.
Ghiotto \etal~\cite{ghiotto-tse18} et al. perform a large-scale study of merge conflicts over Java projects and characterize the nature of merge conflicts and patterns in resolutions. 
The study envisions the need for tools that can capture the patterns in such resolutions.
In addition to complementing the study by focusing on forks, our work can be seen as the first realization of the vision in this paper for a large project by using program synthesis to capture the frequent patterns. 
The problem of resolving merge conflicts soundly has received a lot of attention, going back to the work of program integration~\cite{Horwitz1989}, where static analysis on the three input programs is performed to create a  merge satisfying a conflict-freedom property.
However, such approaches were never implemented or evaluated on real-world benchmarks.
More recently, Sousa \etal~\cite{sousa2018verified} use program verifiers to check for the semantic conflict-freedom notion of correctness of a given merge, but do not synthesize the merge. 
Structured and semi-structured merge tools~\cite{Apel11,lessenich2015} (such as JDIME) resolve some class of textual conflicts by lifting the textual 3-way differencing algorithm to the case of abstract-syntax trees. 
Since we are analyzing C++, we cannot perform a direct comparison due to the lack of such tools for C++.
However, we believe our technique is complementary to such structured algorithms.
One can apply the structured algorithms to first soundly resolve conflicts, and then turn to a synthesis guided approach to fall back on the learned resolution for the remaining. 
In addition, we believe that the structured techniques will not apply to most of the cases we encounter in this paper, due to the asymmetric nature of the fork related changes (e.g. AFSC), and renaming or movement of header files.
Finally, the work of Sung \etal~\cite{sungtowards} studies and proposes fixing build breaks introduced by a merge in main/fork structure but does not deal with the problem of textual merge conflicts. 

\ourpara{Bug Fixing:}
Recently, Bader \etal \cite{bader2019getafix} applied a hierarchical clustering technique to address the fixing bugs by learning through examples. In this study, the AST based differencing technique has been used to identify the edits and clusters the edits based on the patterns. Finally, these fixes are applied to the code to address 7 different patterns in Java, e.g., null pointer exception, boxed primitive constructor, new class instances, etc. However, in our work, we empirically evaluated the merge conflicts in \edge{} and take 1-2 examples for each pattern to build programs that can assist developers in resolving the \texttt{Include} and \texttt{Macro} related conflicts.

\ourpara{Program Synthesis:}
Prior work applied program by examples (PBE), a form of program synthesis whose specification is examples, to various domains~\cite{flashfill, le2014flashextract, meng2011systematic}.
Gulwani introduced FlashFill, a system that synthesizes string transformation scripts from examples~\cite{flashfill}. 
FlashExtract~\cite{le2014flashextract} allows end users to extract hierarchical data from semi-structured files by simply highlighting some examples. 
Anderson and Lawall \cite{andersen2010generic}, Anderson \etal \cite{andersen2012semantic} proposed techniques to learn version update patches for Linux files from input-output examples. 
Meng \etal \cite{meng2011systematic, lase} developed SYDIT and LASE to synthesize code edit scripts from examples. 
Our work targets a new, different domain that has not been explored by PBE: merge conflict resolutions.
Unlike prior work, which usually operates on a single snapshot of a file or program, our synthesizer learns patterns and transformation from various places: changes in the main branch, changes in the forked branch, and also file content outside the conflicting regions.



\section{Threats to Validity}
\label{sec:threats}
\noindent\textit{\textbf{Construct Validity:}}
Replaying merges and identifying conflicts have been done using the git-based system. Given the broad and extensive use of such system, we believe that errors in measurement during the dataset collection are avoided.

\noindent\textit{\textbf{Internal Validity:}} The empirical evaluation, the pattern recognition, and the domain-specific language built from the patterns are done as a part of a case study on \edge. Also, the patterns were recognized and tested on 12 weeks of data. Though we believe that learning from the past~\cite{ghiotto-tse18} can have a potential impact on the resolution of merge conflicts, new patterns can emerge in the future, and that can be addressed in two possible ways—one, learning the new pattern from the existing DSL. Second, update the DSL to accommodate the new patterns. A possible threat is that it is possible that some scenarios are not detected by our empirical evaluation in \S\ref{sec:classification} due to the 12 week timeline. Another possible threat can be overfitting. To remediate the overfitting, we have taken two steps. First, we built our ranking function to choose smaller and more general programs based on the Occam's Razor principle. Simultaneously, the DSL has to be expressive enough to cover as many tasks as possible while being concise enough to be learned efficiently. Keeping both concerns in mind, we consciously bias our DSL to choose the smaller programs by punishing the longer ones using the ranking function. Second, the overfitting can also be removed by applying more examples for the learning process. For this study, we used 2-3 examples for each pattern. Our solution can be extended to incorporate more examples to remove overfitting if needed in the future. 
However, since our approach and the DSL has been based on the patterns and common resolution strategies found in the empirical evaluation, there can be scenarios, where our approach may not work and needs re-training. For example, all the fixing strategies are based on a simple fact that the line(s)/Nodes from either the main or fork section of the conflict will be removed or concatenated. There might be scenarios where none of the strategies are applicable—for instance, taking both the main and fork section of the conflict and changing some parameters. In such scenarios, our approach will not be able to generate any solution. However, this situation can be tackled by extending the underlying implementation of the \texttt{Selection} operations in our DSL. Currently, \texttt{[Main/Fork]ByPath} works for selecting an include path node. Updating these operations by adding the selection of full AST nodes can help to identify such complex patterns.
Also, these patterns can change in the future, e.g., the pattern illustrated in the Frequent Behavior (an example of a project-specific pattern), where due to the different logging module in the main and fork, the main one gets removed. If there is a change in the logging functionality and developers prefer to remove the fork version of the logging, then a new programs need to be generated with new examples.

\noindent\textit{\textbf{External Validity:}}
The current results are based on \edge. Though \edge{} is a widely used system, drawing general conclusions from empirical studies in software engineering is difficult because any process depends to a large degree on a potentially large number of relevant context variables \cite{basili1999family}. For this reason, we cannot assume a priori that the results of a study generalize beyond the specific environment in which it was conducted\cite{basili1999family}. Researchers become more confident in a theory when similar findings emerge in different contexts \cite{basili1999family}. Towards this end, we intend that our case study will help contribute towards building a body of knowledge in the area of using program synthesis for merge conflict resolution by replicating across different context variables and environments.
\section{Conclusion and Future Impact}
\label{sec:conslucion}
In this paper, we investigate the problem of merge conflicts on a large system. Empirically, our findings identify the type of files, size of conflicts, location of conflicts in a file, and resolution patterns. In our case study, we found that a majority of conflicts in \edge{} occur in C++ files, and overall, 28\% of the merge conflicts are 1-2 line changes. We empirically characterize the different types of operations that developers perform while resolving merge conflicts. We also identified that a non-trivial section of these 1-2 liner conflicts is due to the \texttt{Include} and \texttt{Macro} related structural changes. 

 We propose an expressive domain-specific language to identify the structural merge conflicts in \edge. The use of program synthesis provides the flexibility to learning complex patterns from a few examples (1-2 examples). Our study highlights that learning project-specific patterns is beneficial, and we are able to address 11.44\% of merge conflicts in C++ files with 93.2\% accuracy. Applying program synthesis for assisting with merge conflict resolution can be beneficial for two reasons: 1) flexibility to accommodate new patterns, 2) adding explainability to the resulting suggestions.
 
In the future, we plan to extend this study by applying our approach to other large scale projects. We also plan to explore combining program synthesis with machine learning trained on a large dataset to understand if we can deploy within projects without sufficient history, that is learn from one project and apply to another project - cross project learning. 
\IEEEtriggeratref{15}
\bibliographystyle{IEEEtran}
\bibliography{bibliography/ref}

\end{document}